\documentclass{article}

\PassOptionsToPackage{numbers, compress}{natbib}
\usepackage{amsmath}
\usepackage{cancel}
\usepackage{xcolor}
\usepackage{algorithm}
\usepackage{algorithmic}
\usepackage{chngcntr}

\definecolor{mypink1}{rgb}{0.858, 0.188, 0.478}

\definecolor{alexcolor}{rgb}{0.358, 0.588, 0.178}

\usepackage[preprint]{neurips_2026}

\usepackage[utf8]{inputenc} 
\usepackage[T1]{fontenc}    
\usepackage{hyperref}       
\usepackage{url}            
\usepackage{booktabs}       
\usepackage{amsfonts}       
\usepackage{nicefrac}       
\usepackage{microtype}      
\usepackage{xcolor}         
\usepackage{graphicx} 

\title{Toward a mechanistic understanding of inference in visual cortex and diffusion models}

\author{%
  \textbf{Zeyu Yun}\thanks{Corresponding authors: \texttt{chobitstian@berkeley.edu}, \texttt{baolshausen@berkeley.edu}}$^{\ \,,1,4}$ \quad
  \textbf{Alexander Belsten}$^{3,4}$ \quad
  \textbf{Dasheng Bi}$^{1,4}$ \quad
  \textbf{Zahra Kadkhodaie}$^{5}$ \\[1.5ex]
  \textbf{Yubei Chen}$^{6}$ \quad
  \textbf{Bruno A. Olshausen}$^{\ast,2,3,4}$ \\[3ex]
  \normalfont
  $^{1}$ Dept. of Electrical Engineering and Computer Science, UC Berkeley \\[0.5ex]
  $^{2}$ Helen Wills Neuroscience Institute, UC Berkeley \\[0.5ex]
  $^{3}$ Herbert Wertheim School of Optometry and Vision Science, UC Berkeley \\[0.5ex]
  $^{4}$ Redwood Center for Theoretical Neuroscience, UC Berkeley \\[0.5ex]
  $^{5}$ Flatiron Institute \quad
  $^{6}$ Dept. of Electrical and Computer Engineering, UC Davis
}

\begin{document}

\maketitle
\begin{abstract}
We describe a model of perceptual inference in primary visual cortex (V1) equivalent to a minimal diffusion model whose function can be readily understood from its parameters. The model is based on sparse coding with a non-factorial prior over latent variables in the form of an unconstrained, pairwise interaction matrix, extending standard sparse coding inference to a general recurrent dynamical system. We efficiently train these recurrent dynamics using a denoising score-matching objective and implicit differentiation. After training on natural images, the learned interaction matrix mirrors the structure of horizontal connections in superficial layers of V1 that link neurons of similar orientation tuning. This model exhibits exceptionally good denoising performance, restoring image features such as extended contours amid extreme visual ambiguity, nearly matching the behavior of standard, black-box diffusion architectures in generalization regime. Owing to the model's simplicity, the network’s Jacobian can be decomposed directly in terms of the interaction matrix between latent variables, revealing mechanistically how the recurrent dynamics assign high probability over a continuous family of natural structural deformations. Intriguingly, within this circuit, a large fraction of latent variables learn to disconnect from visual input altogether, essentially forming a hierarchical representation that appears to enforce global consistency among image features. Together, the model and results bridge two distinct domains: for neuroscience, it generates concrete, testable hypotheses regarding functional connectivity in recurrent neural circuits during perceptual inference tasks; for machine learning, it elucidates the internal mechanisms learned by diffusion models that allow them to generate infinitely many novel images from a finite training set. 

\end{abstract}

\section{Introduction}

\begin{figure}[t]
    \centering
    \includegraphics[width=\linewidth]{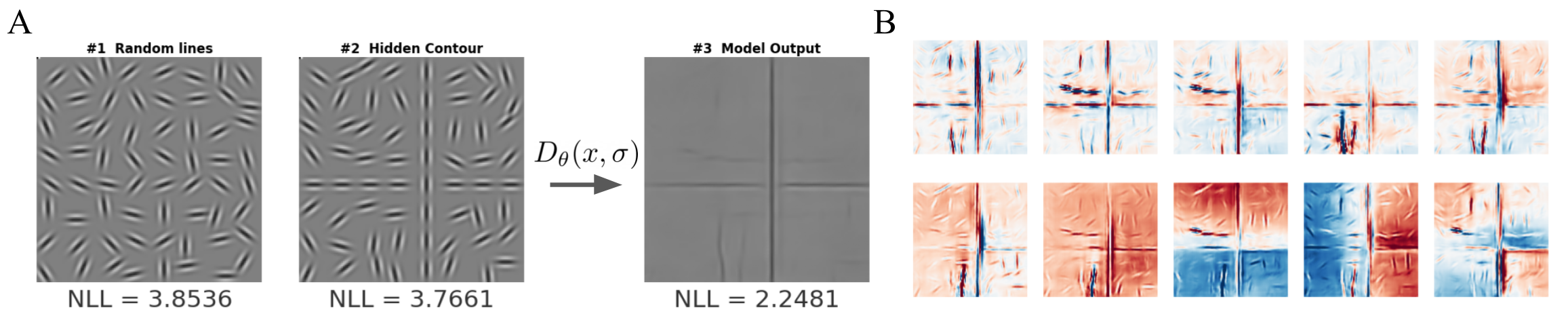}
    \caption{\textbf{Likelihood and Jacobian analysis of contour stimuli.} 
    \textbf{(A)} Estimated negative log-likelihood (NLL), reported in bits/dim (see Appendix \ref{appendix:BPD}), for three input images. Likelihood strictly increases (lower NLL) across the sequence: purely random lines (\#1), random lines with hidden disconnected contours (\#2), and the diffusion model's output when the clean image \#2 is evaluated at a nonzero noise embedding, denoted as $D_\theta(x, \sigma) = x + \sigma^2 \nabla_x \log p_\sigma(x)$ (\#3). 
    \textbf{(B)} Visualization of the principal eigenvectors of the Jacobian $J(x)$ of the denoiser, evaluated at the hidden contour image and ranked by eigenvalue magnitude.}
    \label{fig:intro}
\end{figure}

When the image of Figure~\ref{fig:intro}A (middle) is viewed by human observers, it is immediately and effortlessly perceived as containing a horizontal and vertical contour, whereas an image composed of the same Gabor elements randomly arranged is seen as structureless~\cite{field1993contour}.  Interestingly, a diffusion model trained on natural images perceives these Gabor images similarly, as revealed by its output (Figure~\ref{fig:intro}A, right). In particular, the
principal eigenvectors of the model's Jacobian reveal the perturbations that induce high-probability variants of the input in image space (Figure~\ref{fig:intro}B): in this case, spreading the discrete Gabor elements into a continuous contour. The question we address here is, why is this happening -- both in visual cortex and diffusion models -- and what specific neural computational mechanisms are needed to capture this in a model of perception?

While interrogation of a diffusion model reveals its probability gradient for specific images \cite{mohan2019robust}, such analysis is only local and anecdotal, and we still lack an explanation for why this arises from the specific mechanisms within the neural network of the model. On the other hand, by studying visual cortex of the brain, we can identify specific neuroanatomical structures that suggest how such contour completion occurs -- namely, the network of horizontal connections in superficial layers that spread activation among neurons with spatially separate but similarly oriented (and in some cases co-linearly aligned) receptive fields~\cite{hirsch1991synaptic,fitzpatrick1996functional,field1993contour,garrigues2007learning,lee2003hierarchical}.
It has long been theorized that this connectivity could underlie perceptual grouping phenomena such as contour integration, whereby geometrically aligned features are perceived as belonging to a common object~\cite{hirsch1991synaptic,field1993contour}. This is also consistent with the evidence that neural dynamics perform sampling-based probabilistic inference \cite{orban2016neural, echeveste2020cortical}.
Although numerous models have been proposed~\cite{parent1989trace,grossberg1997visual,li1998neural, george2025detailed, toosi2025generative}
, it has yet to be shown how this connectivity structure could emerge from a probabilistic, generative model trained to perform inference on natural images (a notable exception is~\cite{george2017generative} which trained a probabilistic graphical model on images of line drawings).  

Here we approach this problem by building on the \emph{sparse coding model} of V1, which has been proposed to account for its oriented, simple-cell receptive fields in terms of a linear generative model of natural images with a sparse, factorial prior over latent variables~\cite{olshausen1996emergence,olshausen1997sparse, field1987relations,field1994goal,olshausen2014natural}. Inspired by horizontal connections in visual cortex, we add to the sparse coding model an interaction matrix between latent variables with the goal of learning the co-occurrence statistics of features at different spatial locations, orientations and spatial scales \cite{simoncelli1999modeling, portilla2000parametric,hyvarinen2009natural,lee2003nonlinear, carlsson2008local, chen2018sparse, chen2022bag}. 
Previous theoretical and modeling work has shown that co-circularity of oriented elements constitutes an important aspect of natural scene statistics~\cite{parent1989trace,geisler2001edge,sigman2001common}.  Here we reason that by training the model on natural images, it should be possible to incorporate these statistics into the prior, which should in turn improve the ability to resolve ambiguity in images.

In what follows, we first provide relevant background and then formally specify the proposed model. We then show that the learned interaction matrix weights are aligned with co-linear facilitation, and that the model's performance on denoising tasks is far superior to standard sparse coding and competitive with deep diffusion models. Local linear analysis of the model's behavior via its Jacobian reveals how activity spreads among latent variables in a context-dependent manner, leading to the completion of contours amidst ambiguity. Crucially, this recurrent spreading mechanism also explains how diffusion models assign high probability to realistic structural deformations, illuminating the internal mechanism that allows them to sample novel images.

\section{Background and preliminaries}
{\bf Sparse coding model of primary visual cortex.} The sparse coding model posits that neural activity represents latent variables, $z$, that encode an image, $x$, by descending an 
%
%
energy function $E(x,z) =
    \frac{1}{2\sigma^2}\|x-\Phi z\|_2^2
    + \lambda ||z||_{1}$.
This energy function balances stimulus reconstruction error with neural sparsity. From a Bayesian perspective, the first and the second terms correspond to the likelihood and the prior respectively. When trained on natural images to maximize the log-likelihood of the model, $p_\theta(x)\propto\int e^{-E(x,z)}dz$, the dictionary ($\Phi$) converges to a set of Gabor-like functions similar to V1 receptive fields \cite{olshausen1997sparse}.
Typically, this optimization  process relies on alternating between two objectives, where the sparse latent codes are inferred using algorithms like Iterative Shrinkage-Thresholding (ISTA) \cite{ISTA}, and the dictionary learned from data.
However the underlying assumption of statistical independence in the prior (corresponding to $||z||_{1}$) is a severe limitation, as
%
%
natural image features are not statistically independent: oriented features exhibit strong dependencies across position, scale, and orientation~\cite{hyvarinen2000emergence,portilla2000parametric,schwartz2001natural, hyvarinen2001topographic, karklin2003learning, karklin2005hierarchical}. These dependencies are central to perceptual phenomena such as contour integration and completion, where spatially separated but geometrically aligned elements are perceived as grouped together. 
%
%
Several extensions have been proposed to model non-factorial structure in the latent variables. Hierarchical sparse coding models introduce additional latent variables that capture dependencies among lower-level features~\cite{karklin2005hierarchical,yu2011learning,hosoya2015hierarchical,hosoya2016learning,boutin2020effect}. Other single-layer models impose structured priors, such as topographic or group sparsity, to encourage related features to co-activate~\cite{hyvarinen2000emergence,paiton2020subspace}. These approaches capture some non-factorial structure, but often rely on rigid, hand-specified priors that cannot be learned. More flexible non-factorial models can learn dependencies between latent variables~\cite{garrigues2007learning,garrigues2010group, olshausen2007bilinear, osindero2006topographic}, but their inference and learning procedures are typically more difficult and may require sampling over high-dimensional latent variables.

{\bf Training and sampling of Denoising diffusion models.} In recent years, Denoising Diffusion Models (DDMs) have emerged as an effective way to learn densities of natural images from finite datasets \cite{vincent2008extracting, sohl2015deep, saremi2019neural, ho2020denoising,Song+21SDE}. 
At their core, these models learn the score function—the vector field pointing towards regions of higher data probability. In practice, the true score is approximated by a deep neural network $s_\theta(x, \sigma)$ trained to minimize the Denoising Score Matching (DSM) loss: 
\begin{equation}
\label{eq:DSM} \mathcal{L}_{\mathrm{DSM}} = \mathbb{E}_{\sigma, y, \epsilon} \left[ w(\sigma) \left\| s_\theta(x, \sigma) + \frac{\epsilon}{\sigma} \right\|_2^2 \right], \quad \text{where} \quad x = y + \sigma \epsilon 
\end{equation}
where the clean image is $y$, the noisy image is $x = y + \sigma \epsilon$ with $\epsilon \sim \mathcal{N}(0, I)$, and $w(\sigma)$ is a positive weighting schedule. After training, $s_\theta(x, \sigma) \approx \nabla_x\log p(x)$. 
While empirically powerful, $s_\theta$ is typically parameterized by a black-box neural network, 
making it difficult to understand how individual or populations of neurons structurally model the image prior. In this work, we leverage the powerful denoising score matching objective, to learn the parameters of our non-factorial sparse coding model. The explicit nature of the model allows for direct analysis of its neural circuit, shedding light on how diffusion models mechanistically assign high likelihood to natural images.


{\bf Understanding feature learning in generalized diffusion models. }
With a sufficiently expressive function, $s_\theta$, the optimal minimizer of Equation \ref{eq:DSM} is the empirical density which results in memorization of the training set. Nevertheless, it has been shown that with sufficient data, diffusion models do not memorize and generate novel samples from the underlying distribution, thanks to inductive biases of the architecture and training \cite{kadkhodaie2023generalization, kamb2024analytic, niedoba2024towards, liang2024diffusion}. Once in this generalization regime, a natural question is \emph{what features did the model extract from data during training?} One approach to answering this question is local linear analysis through eigen-decomposition of the Jacobian of the output w.r.t the input of the denoiser \cite{mohan2019robust, kadkhodaie2023generalization}. By Tweedie’s formula, this Jacobian is analytically linked to the Hessian of the learned log-likelihood ($J(x) = I + \sigma^2 \nabla_x^2 \log p_\sigma(x)$), meaning its dominant eigenvectors point directly toward the high-probability states defined by the model. Empirically, these directions correspond to naturalistic structural deformations—specifically, oscillatory bases tied to image contours—which constitute the exact features that allows the model to generalize \cite{kadkhodaie2023generalization}. While these analyses have revealed that the model learns these features, they come short in offering a mechanistic explanation of how these features arise from the activities of individual neurons in the network. The simplicity of our model provides the opportunity to expand such Jacobian analyses to shed light on the mechanisms that enables generalization at a neuronal level. 

\section{Non-factorial sparse coding as a minimal diffusion model}

We now define a sparse coding model that learns a non-factorial prior over the latent variables in the form of an unconstrained, pairwise interaction matrix. This augmented prior allows the model to capture dependencies among oriented features across positions, scales, and orientations.
The specific non-factorial prior we assume takes the form $z^\top Mz$, where the interaction matrix, $M$, captures pairwise dependencies between the latent variables. 
The resulting energy function is
\begin{equation}
E_{\theta,\sigma}(x,z) = \tfrac{1}{2 \sigma^2}\|x - \Phi z\|_2^2 + \gamma_{\psi}(\sigma) \big( \|\lambda \circ z\|_1 + \tfrac{1}{2} z^\top M z \big).
\label{eq:joint_energy}
\end{equation}
As in the original sparse coding model, $\Phi$ is a dictionary of image features, while latent variables $z$ represent neural activity, which is regularized toward sparsity via an $L_1$ penalty. 
Each element $z_i$ is associated with a learned sparsity parameter $\lambda_i$, and $\circ$ denotes the Hadamard (element-wise) product. To handle varying levels of image corruption, the scalar function $\gamma(\sigma)$ dynamically learns the optimal balance between the data likelihood and the augmented prior. This function is parameterized by a two-layer MLP with weights $\psi$, which processes $\sigma$ via sinusoidal Fourier features. We define the total set of trainable parameters as $\theta = \{\Phi, M, \lambda, \psi\}$.

\subsection{Training non-factorial sparse coding model via denoising score matching}
Sparse coding with flexible non-factorial prior is notoriously difficult to train with standard maximum likelihood estimation. To overcome this, we train the model using Denoising Score Matching (DSM). Because DSM relies on matching the model's score function to that of the noisy image, our first step is to derive the explicit score function for non-factorial sparse coding model. 

Using the energy function of Equation \ref{eq:joint_energy}, marginalizing over the latent variable $z$ results in the data density,  $p_{\theta,\sigma}(x) \propto \int \exp(-E_{\theta,\sigma}(x,z)) dz$. This integration, however, is in high dimensions and thus computationally intractable. Hence, we settle for an approximate solution using a MAP approximation. By assuming the posterior, $p_{\theta,\sigma}(z|x)$, is highly concentrated around its mode, the integral becomes overwhelmingly dominated by its peak. Therefore, we can bypass the full integration and simply obtain the score via the energy-minimizing latent code: $\nabla_x \log p_{\theta,\sigma}(x) \approx -\nabla_x E_{\theta,\sigma}(x, z^*(x;\sigma,\theta))$, where $z^*(x;\sigma,\theta) = \arg\min_z E_{\theta,\sigma}(x,z)$ is the MAP estimate. Note that $x$ only appears in the reconstruction term of the energy function, so the approximate score simplifies to a closed-form expression of the optimal latent state:
\begin{equation}
\nabla_x \log p_{\theta,\sigma}(x) \approx \tfrac{1}{\sigma^2}(\Phi z^*(x;\sigma,\theta) - x)
    \label{eq:fix-point_approximate}
\end{equation}
Now, we arrived at a simple and explicit parameterization of the score in which the dictionary $\Phi$ and the optimal latent variable $z^*$ should be learned. Substituting Equation \ref{eq:fix-point_approximate} back into the Denoising Score Matching objective (Equation \ref{eq:DSM})  yields a simple, weighted reconstruction objective. Refer to Appendix \ref{app:score_matching} for derivation and the exact noise weighting $w(\sigma)$ used for training: 
\begin{equation}
    \mathcal{L}_{\mathrm{DSM}}(\theta) = \mathbb{E}_{\sigma, y, \epsilon}\left[ w(\sigma) \left\| \Phi z^*(x;\sigma,\theta) - y \right\|_2^2 \right], \quad \text{where} \quad x = y + \sigma \epsilon 
    \label{eq:dsm_to_weighted_recon_sigma4}
\end{equation}


In summary, training non-factorial sparse coding model reduces to training a dynamical system to denoise: for a perturbed observation $x = y + \sigma \epsilon$, we first infer the MAP latent variables $z^*$, and then update the model parameters to minimize the reconstruction error in Equation~\ref{eq:dsm_to_weighted_recon_sigma4}. 
Similar to classic sparse coding, this learning procedure alternates between inferring latent coefficients and updating the dictionary.

\subsection{MAP-Induced Recurrent Inference}
We use the iterative shrinkage-thresholding algorithm (ISTA)~\cite{beck2009fast, rozell2008sparse} to compute the MAP latent variables $z^*(x;\sigma,\theta)$ for a given input.
By taking a gradient step on the differentiable components of the energy (the reconstruction error and quadratic coupling) and applying the proximal operator to the non-smooth prior ($L_1$ penalty and non-negativity constraint), the optimization unfolds as the dynamics of a recurrent neural circuit. 
Starting from $z_0=0$, the latent variables are updated until convergence to the fixed-point $z^*$:
\begin{align}
    z_{t+1} = T(z_t;\sigma) =\text{ReLU} \left( z_t + \eta \left[ \underbrace{\Phi^\top x}_{\text{Feedforward Drive}} - \underbrace{\left( \Phi^\top \Phi + \sigma^2\gamma(\sigma) M \right) z_t}_{\text{Recurrent Drive}} - \underbrace{\gamma(\sigma) \lambda}_{\text{Sparsity Bias}} \right] \right)
    \label{eq:ff_rec_update}
\end{align}
given a step size $\eta > 0$.
This update rule can be separated into three terms.
A feedforward input drive equal to the projection of the noisy input $x$ onto the dictionary $\Phi$.
A recurrent drive  which is computed from the current state of the network $z_t$. The recurrent drive has two components: a Gram matrix $\Phi^\top \Phi$ which arises from the likelihood, and a noise-weighted $\sigma^2\gamma(\sigma) M$ that favors or suppresses co-active latent variables according to the non-factorial prior.
Lastly, the sparsity bias $\gamma(\sigma) \lambda$ subtracts a fixed value before the ReLU.



\subsection{Differentiation through the dynamics with implicit function theorem}
Both training the denoising objective and analyzing the model via the Jacobian $J(x)$ require differentiating through the convergent state $z^*(x;\sigma,\theta)$. Rather than unrolling the entire iterative trajectory of Equation~\ref{eq:ff_rec_update} via Backpropagation Through Time (BPTT), which incurs prohibitive memory and computational costs, we exploit the fact that $z^*$ converges to a stable fixed point.
By defining the fixed-point residual $F(z^*; x)=z^*-T(z^*;x)=0$ 
, the Implicit Function Theorem (IFT) allows us to compute the exact analytical gradients in a single, closed-form step:
\begin{equation}
        \tfrac{\partial z^*}{\partial \theta}=-\left[\nabla_z F\right]^{-1}\nabla_\theta F, 
    \qquad 
    \tfrac{\partial z^*}{\partial x}=-\left[\nabla_z F\right]^{-1}\nabla_x F\label{eq:ift_gradients}
\end{equation}

The first gradient $\frac{\partial z^*}{\partial \theta}$ enables efficient parameter updates during training via the chain rule. The second gradient $\frac{\partial z^*}{\partial x}$ is the link for our subsequent analysis: because our explicit model formulates the denoised output as a linear synthesis of the latent state ($f_\theta(x)=\Phi z^*$), applying the chain rule directly yields the exact analytical denoising Jacobian $J(x)=\Phi \frac{\partial z^*}{\partial x}$ (see Appendix~\ref{app:implicit gradient}). 

The approach of training recurrent neural networks which arrive to fixed points using the IFT was originally described by Pineda~\cite{pineda1987generalization} and Aleimda~\cite{almeida1990learning}. Recent work has applied this method to deep learning architectures~\cite{bai2019deep, geng2021training, el2021implicit,bai2024fixed,wang2025hierarchical} to achieve impressive results. We use recent improvements to this algorithm to efficiently estimate the gradient~\cite{fung2022jfb} without computing the inverse of $\nabla_zF$, we use un-rolling based phantom gradient to estimate the implicit gradient \cite{geng2021training} (details in Appendix \ref{app:appendix_ift}).

\subsection{Mechanistically Decomposing the Jacobian} \label{sec:mech_decomp}
Due to the simplicity of the model, we can further open the black by decomposing the denoising Jacobian $J(x)$. Combining all the recurrent drive into a single matrix $W = \Phi^\top \Phi + \sigma^2\gamma(\sigma) M $, we can expand the implicit gradient in Equation~\eqref{eq:ift_gradients} to decompose the denoising Jacobian in the following form:
\begin{equation}
J(x) = \frac{d x_{\text{clean}}}{d x_{\text{noise}}} = \Phi \underbrace{\eta \left[ I - \Sigma(I - \eta W) \right]^{-1} \Sigma}_{J_z} \Phi^\top = \Phi J_z \Phi^\top
\label{eq:jacobian-change-basis}
\end{equation}
where $\Sigma = \text{diag}(1_{z^* > 0})$ acts as a binary gating matrix representing the currently active sparse code. This formulation explicitly reveals a structural change of basis: the pixel-space Jacobian $J(x)$ is the Jacobian in latent-space $J_z$ projected through the dictionary $\Phi$. We can further expand the matrix inverse into an infinite Neumann series ($[I - A]^{-1} = \sum_{k=0}^\infty A^k$), as $J_z = \sum_{k=0}^\infty (\eta \Sigma W)^k \Sigma$. 

To understand $J_z$, we examine a single column $J_{z,i} = J_z e_i$, which represents the network's impulse response to perturbing the $i$-th latent variable. Because the binary diagonal gating matrix $\Sigma$ only passes active neurons ($\Sigma e_i = e_i$ if $z^*_i > 0$, and $0$ otherwise), any perturbation to an inactive neuron is immediately silenced. For an active neuron, we can use the Neumann series expansion to decompose its impulse response in the following highly interpretable manner:
\begin{equation}
    J_{z,i} = \underbrace{e_i}_{\text{Sparse gating}} + \underbrace{\eta \Sigma W e_i}_{\text{1st-order spread}} + \underbrace{\eta^2 \Sigma W \Sigma W e_i}_{\text{2nd-order spread}} + \dots \quad (\text{if } z^*_i > 0, \text{ else } 0)
    \label{eq:jacobian-decomposition}
\end{equation}

Together with the change-of-basis in Equation~\ref{eq:jacobian-change-basis}, this formulation provides a transparent, mechanistic explanation for how the network processes incoming perturbations.
First, perbutation in pixel is projected from and to latent space by dictionary $\Phi$. The $\Sigma$ matrix then acts as a mask that silences perturbation on inactive neurons, ensuring only contextually relevant signals survive. Next, during \textbf{context-aware spread} ($\eta \Sigma W \Sigma$), these surviving signals propagate via lateral connections $W$. The outer $\Sigma$ strictly clamps this activity spreading, ensuring cross-talk between neurons only leak into currently activated neurons. Furthermore, the infinite tail of the sum $(\eta \Sigma W)^k$ represents higher-order, continuous spreading of neural activity within the context.

\section{Results}
\subsection{Emergence of long-range horizontal connections}
\label{result:horizontal connections}

\begin{figure}[h]
    \centering
    \includegraphics[width=1.0\linewidth]{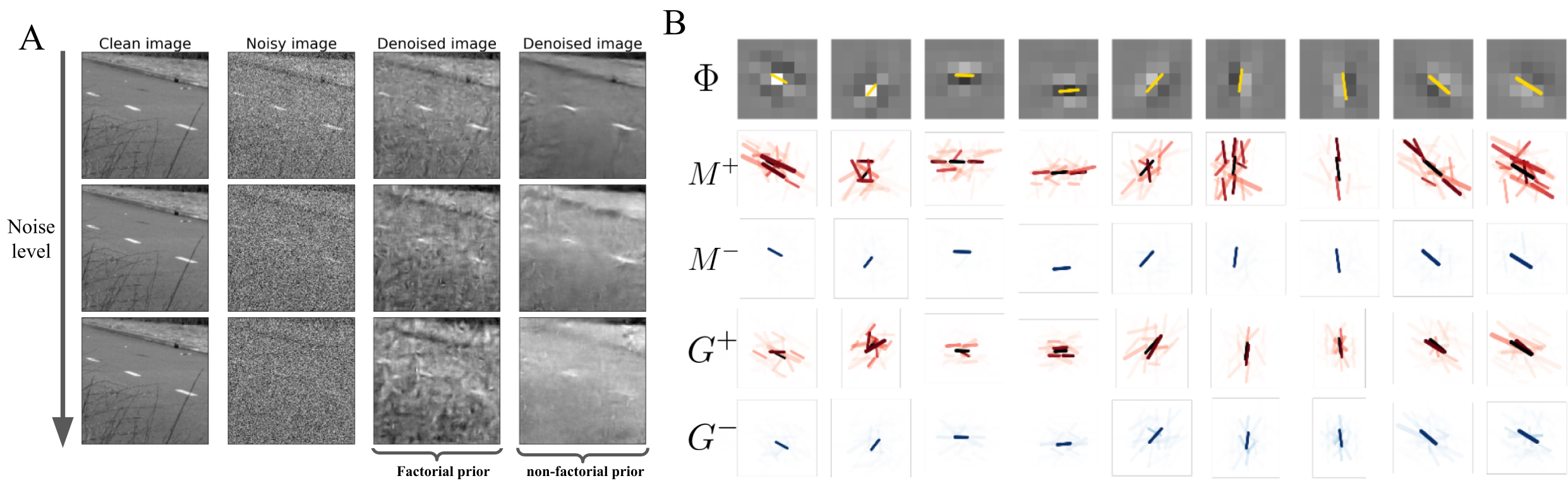}
    \caption{
        \textbf{Emergence of V1-like connectivity and denoising performance.} 
        \textbf{(A)} 
        Denoising comparison across increasing noise levels (rows) showing clean, noisy, factorial, and non-factorial reconstructions. 
        \textbf{(B)} 
        Visualization of learned latent interactions for a subset of dictionary elements. 
        Each column is centered on one reference basis function $\Phi_i$ shown in the top row. The yellow/black bar indicates its position and preferred orientation. 
        Rows below show needle plots of its interactions with other basis functions $\Phi_j$: excitatory and inhibitory components of the learned pairwise matrix $M_{ij}$, followed by excitatory and inhibitory components of the Gram matrix $G_{ij}=\Phi_i^\top \Phi_j$. 
        Each colored bar is placed at the spatial location and orientation of $\Phi_j$, and color saturation indicates interaction magnitude. 
    }
    \label{fig:neuroscience}
\end{figure}

When trained on a dataset of luminance-only natural scenes (i.e gray-scale images)~\cite{van1998independent}, the model learns a dictionary $\Phi$ of localized, oriented Gabor-like basis functions, mirroring V1 simple cell receptive fields (Figure~\ref{fig:neuroscience}B). The inference dynamics (Equation~\ref{eq:ff_rec_update}) decompose the lateral interactions into two mechanisms: the Gram matrix ($\Phi^\top\Phi$), and the learned non-factorial structural prior ($M$). As visualized in Figure~\ref{fig:neuroscience}B, $M$ learns \textit{collinear facilitation}. It develops strong excitatory connections between spatially offset neurons whose preferred orientations broadly co-align along a shared axis, alongside strong self-inhibition to enforce sparsity. 
This emergent collinear connectivity  parallels the `association field' and co-facilitation principles proposed in~\cite{field1993contour}. Computationally, $M$ acts as the mechanism to bind local edge segments into extended, continuous contours.
By contrast, the Gram matrix exclusively mediates more localized interactions between neurons with overlapping receptive fields (quantified in Figure~\ref{fig:h_conn_summary}). 

To understand the practical gains of learning a nonfactorial prior relative to a factorial model, we trained another model which has fixed $M_{ij}=0\ \forall ij$, but is otherwise identical. As shown in Figure~\ref{fig:neuroscience}A, we find that the non-factorial model performs significantly better denoising than its factorial counterpart. Specifically, the factorial model hallucinates dense, spurious edges in noisy, unstructured regions because it evaluates features in isolation. Conversely, our non-factorial prior leverages $M$ to explicitly enforce the continuity of true structural edges while actively suppressing uncoordinated background noise.

\subsection{Mechanistic decomposition of implicit prior in diffusion model}
\label{result:generalization}

\begin{figure}[h]
    \centering
\includegraphics[width=1.0 \linewidth]{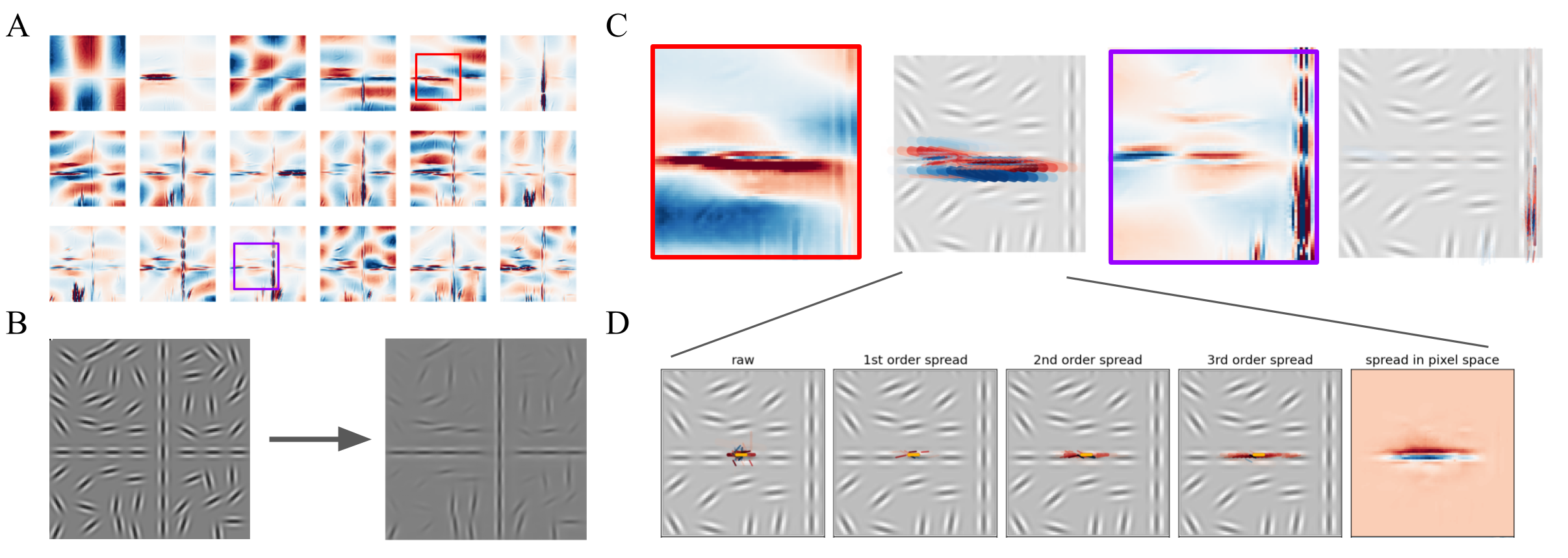}
    \caption{{\bf Mechanistic decomposition of the denoising Jacobian.} (A) Dominant eigenvectors of the pixel-space Jacobian $J(x)$ for the non-factorial sparse coding model ($\sigma = 0.17$). Boxes denote zoom regions for panel C. (B) The model's denoised output $D(x,\sigma)$ on a hidden contour stimulus. (C) Paired pixel-space and latent-space eigenvectors from the highlighted regions. Latent activations are shown as needle plots (orientation denotes the neuron's preferred receptive field; color denotes activation intensity), revealing that participating neurons strictly align with the completed contour. (D) Unrolling a single column of the latent Jacobian (Equation \ref{eq:jacobian-decomposition}). Perturbing a single active neuron (gold needle) yields unconstrained raw spread ($W e_i$) with off-axis leakage. Subsequent columns show higher-order recurrent terms ($(\Sigma W)^n e_i$), demonstrating how sparse gating strictly confines the excitatory (red) and inhibitory (blue) spread to the active collinear context, culminating in the fully integrated impulse response projected into pixel space. (see Figure ~\ref{fig:more_eigenvis}~\ref{fig:more_jacobian_spread} for visualization of more basis.)}
    \label{fig:harmonic}
\end{figure}

As shown in the background section, the implicit prior of a diffusion model is revealed by the denoising Jacobian, where its dominant eigenvectors correspond to the high probability directions resulting from this prior. Figures~\ref{fig:intro}B and \ref{fig:harmonic}A demonstrate that these eigenvectors manifest as oscillatory bases inherently tied to the contours of the input images in both standard U-Nets and our model. 
Empirically, these directions correspond to fully connected contours that bridge the gaps in fragmented inputs, which constitute the exact features the model uses to resolve visual ambiguity.
However, in standard diffusion models, exactly how the neural network constructs these highly structured eigenvectors remains a black box.

Thanks to the architectural simplicity of our model, we can mechanistically explain exactly how these bases are constructed. To uncover this mechanism, Equation \ref{eq:jacobian-change-basis} allows us to analyze these pixel-space eigenvectors directly within the latent space via a change of basis using the learned dictionary ($\Phi$).\footnote{Projecting pixel eigenvectors via $\Phi^\top$ bypasses the prohibitive cost of computing the full latent Jacobian. This proxy mathematically assumes $\Phi$ forms a tight frame ($\Phi^\top \Phi \propto I$) and serves as an effective approximation for visualization.} Visualizing these projected eigenvectors reveals a fundamental property: these completed contours are coded by neurons that are not only spatially located on the contour, but whose preferred orientations strictly align with it. To understand the global eigenvectors of the latent Jacobian $J_z$, we can start by understanding its individual columns. This is equivalent to decomposing $J_z$ into local impulse responses to track how the network transforms a single active neuron (Equation~\ref{eq:jacobian-decomposition}). This transformation is simply the iterative lateral spreading of activity, explicitly gated by the currently active neurons ($\Sigma$). As shown in Figure~\ref{fig:harmonic}D, if an active neuron spreads its signal unconstrained ($W e_i$), the activation exhibits a strong collinear bias but still suffers from diffuse off-axis leakage. However by applying this learned collinear spread and sparse gating ($\Sigma W e_i$) in an alternating order, the signal can be confined to the set of active neurons. As formalized in Equation~\ref{eq:jacobian-decomposition}, an entire column of the Jacobian is precisely the infinite accumulation of this recurrent spread, constrained to the active support (visualized in Figure~\ref{fig:harmonic}.D). 

Because every local impulse response is restricted to propagate exclusively along the active contour, their macroscopic integration inherently produces eigenvectors that are strictly tied to this geometry. Crucially, this spatially bounded recurrence naturally gives rise to the emergent oscillatory bases observed in Figure~\ref{fig:harmonic}A. In the context of generative modeling, these oscillatory bases are computationally powerful because they enable smooth, continuous geometric interpolations. By linearly combining these bases, the model can locally deform a contour—shifting or bending its structure—while preserving its underlying integrity. Furthermore, because these eigenvectors span the high-probability directions of the learned prior, any linear combination within this subspace maps to another highly likely state, allowing the model to assign high probability to infinite, continuous family of structurally valid deformations. As we explore in the following section, this capacity for modeling continuous, naturalistic deformation forms the exact mathematical foundation of generalization. 

\subsection{Beyond contour completion: Generalization and Emergent Hierarchy}
\label{result: high level generalization}

\begin{figure}[h]
    \centering
    \includegraphics[width=\linewidth]{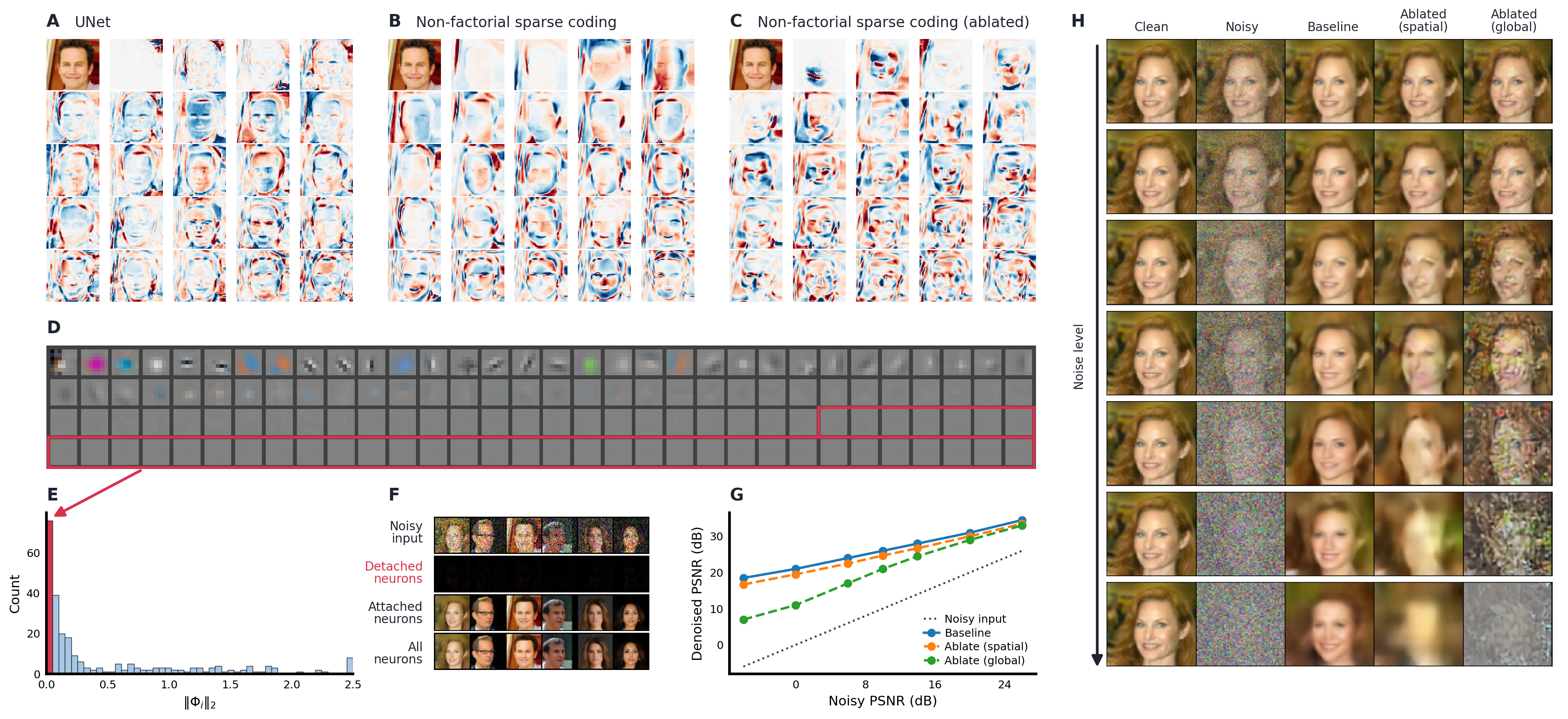}
    \caption{{\bf Semantic generalization and the emergence of detached neurons.} (A-B) Dominant eigenvectors of the pixel-space Jacobian $J(x)$ for a standard U-Net (A) and our non-factorial sparse coding model (B), showing coordinated semantic deformations. (D) Learned dictionary elements ($\Phi$) sorted by norm magnitude. (E) Dictionary norm histogram revealing a structural bifurcation: $\sim$30\% of the population learns near-zero weights, forming functionally decoupled "detached neurons." (F) Sub-population reconstructions: detached neurons yield no visible features, whereas attached neurons capture local geometry. (G) Denoising PSNR drops catastrophically at high noise when detached neurons are ablated (replaced by spatial or global means). (H) Qualitative ablation results across increasing noise levels. 
    }
    \label{fig:face}
\end{figure}

The dominant eigenvectors of the denoising Jacobian reveal not only the model's prior for contour completion, but also its fundamental mechanism for generalization. 
While a finite training set only captures sparse examples of how an object in natural scene might vary, these eigenvectors form a continuous basis for naturalistic deformation. By linearly combining them, the model can synthesize an infinite family of structurally valid configurations that never appeared in the training data. This ability to interpolate and assign high probability to naturalistically deformed configurations is the essence of generalization.
An excellent example of such deformation can be seen in the face dataset (CelebA): the eigenvectors of the denoising Jacobian yield a set of perturbation directions that seamlessly transition the input into a different, yet structurally coherent, face (Figure~\ref{fig:face}A,B).

What is even more intriguing is that when trained on highly structured datasets like the face dataset, both standard U-Nets and our minimal model exhibit semantic-level generalization. While the mechanisms detailed in Section \ref{result:generalization} explain how the recurrent neural circuit constructs the local oscillatory bases responsible for these naturalistic deformations, they do not inherently explain global coherence. Instead of independent, disconnected local deformations, the learned eigenvectors manifest as globally coordinated deformations—such as synchronously shifting both eyes—ensuring all modifications remain structurally consistent with the high-level concept of a "face" (Figure~\ref{fig:face}A). This raises the question: how does a single-layer circuit coordinates these local deformations across the entire image to enforce global constraints?

To understand how a single-layer sparse coding model captures these rules, we examined the magnitude of the learned dictionary elements ($\Phi$) and found that over 30\% of the neural population learns near-zero norm weights (Figure~\ref{fig:face}B-C). We term these \textit{detached neurons} 
because they are functionally decoupled from the ambient pixel space.  
Reconstructing the image exclusively from their activity yields no visible features (Figure~\ref{fig:face}D).

However, these neurons are computationally useful. Ablating their recurrent interactions causes a significant drop in denoising performance, particularly at extreme noise levels ($\text{SNR} \ll 0$) where the network must hallucinate missing structures using global priors (Figure~\ref{fig:face}E-F). Furthermore, without these detached neurons, the global eigenvectors remain tied to the contours of the input image, but they lose the coordinated, long-range structure required to maintain the semantic integrity of a face (Figure~\ref{fig:face}A, right). This suggests a functional organization in which the network uses detached neurons to support hierarchical computation in an otherwise single-layer model. 
By actively decoupling from sensory input, these detached neurons operate as an emergent latent ``second layer," utilizing lateral horizontal connections ($M$) to mediate long-range semantic dependencies and constrain low-level geometric features into globally consistent concepts.

\subsection{Hypotheses for surround excitation in V1}
\label{V1 Hypotheses}

\begin{figure}[h]
    \centering
\includegraphics[width=\linewidth]{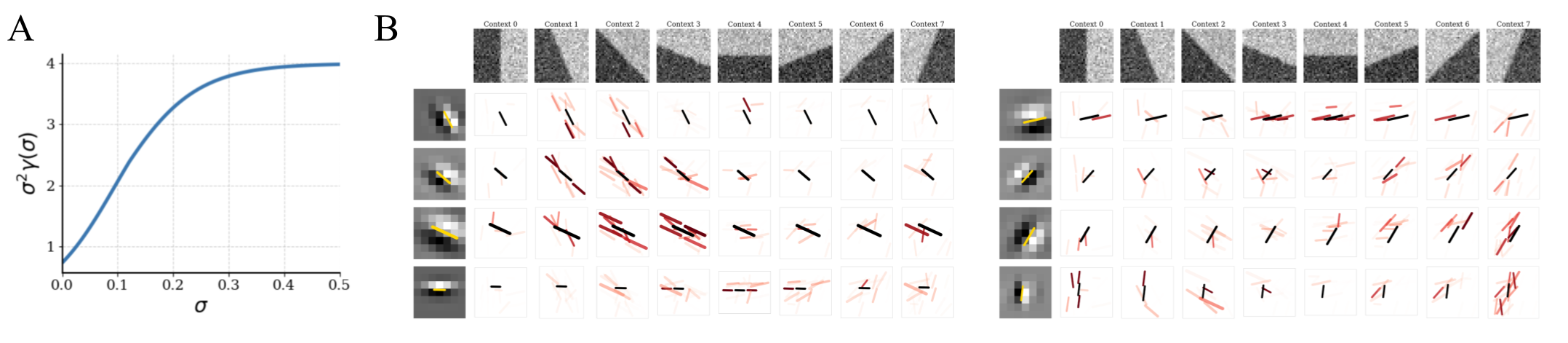}
    \caption{{\bf State-dependent excitation and dynamic geometric routing.} (A) The learned scaling factor $\sigma^2\gamma(\sigma)$ that balances prior vs likelihood contribution in the network's recurrent drive ($\Phi^\top \Phi + \sigma^2\gamma(\sigma) M$). As visual ambiguity ($\sigma$) increases, the network upweights the prior $M$, relying more heavily on learned co-occurrence statistics. (B) Context-dependent excitatory spread for individual target neurons across varying visual stimuli. The top row displays eight noisy line contexts overlapping the target receptive fields. Subsequent rows visualize the target neurons' (black needle) excitatory spread for neighboring neurons (red needle), gated by $\Sigma$. When context is misaligned with a neuron's preferred orientation, the collinear spread is suppressed.
    }
    \label{fig:new_theory}
\end{figure}

Beyond elucidating the mechanisms of generalization in artificial diffusion models, our framework offers testable hypotheses for primary visual cortex (V1). The model predicts how activity propagates in a biological neural circuit in the presence of certain visual stimuli. These predictions are formed by perturbing latent variables (details in Section~\ref{sec:mech_decomp}) and observing the spread of activity. In the model, the network's response to a single-neuron perturbation is dynamically gated by the currently active neural population (Equation~\ref{eq:jacobian-decomposition}). When a model neuron is strongly activated by visual stimulus, perturbing the neuron results in collinear spread (Figure~\ref{fig:new_theory}B). This contrasts with orientation-unspecific spatial spread without perturbation (Figure~\ref{fig:intro}B). This stimulus-dependent routing aligns with recent \textit{in vivo} optogenetic findings, which demonstrate that perturbing a V1 neuron during high-contrast stimulation suppresses nearby neurons most strongly when the stimulus aligns with the target's preferred orientation \cite{Chettih2019}. However, perturbation of the model suggests a new hypothesis: when the stimulus is low-contrast or noisy, the circuit behaves fundamentally differently.

Specifically, in the network dynamics (Equation~\ref{eq:ff_rec_update}), the learned interaction matrix $M$ is scaled by $\sigma^2\gamma(\sigma)$. As visual ambiguity increases, the network transitions from relying on the likelihood term $\Phi^\top \Phi$ to relying on $M$ (Figure~\ref{fig:new_theory}A). Perturbing a model neuron results in excitatory, collinear surround modulation. There is physiological evidence for this noise-dependent shift in V1, where low-contrast stimuli flip the influence of the surround from suppressive to facilitatory \cite{Angelucci2017-fv}. Crucially, the model predicts that this facilitation is not uniform. As visualized in Figure~\ref{fig:new_theory}B, when the target neuron's preferred orientation aligns with the stimulus context, the sparse gate $\Sigma$ restricts the spread of neural activation to a collinear path. Misaligned contexts partially shut these primary gates to allow only weak spread, while orthogonal contexts completely extinguish the lateral signal. Ultimately, we hypothesize that this dynamic gating implements a context-dependent contour completion mechanism: it routes excitatory signals along collinear paths to complete ambiguous or noisy contours (e.g. Figure~\ref{fig:intro}A), while withholding such facilitation when contextual evidence is orientation-incongruent. This suggests dynamics for how the visual system resolves low-level ambiguities, and provides high-level hypotheses for future physiological experiments.

\subsection{Generative Parity with Deep Learning Architectures}
\begin{figure}[htbp]
    \centering
\includegraphics[width=\linewidth]{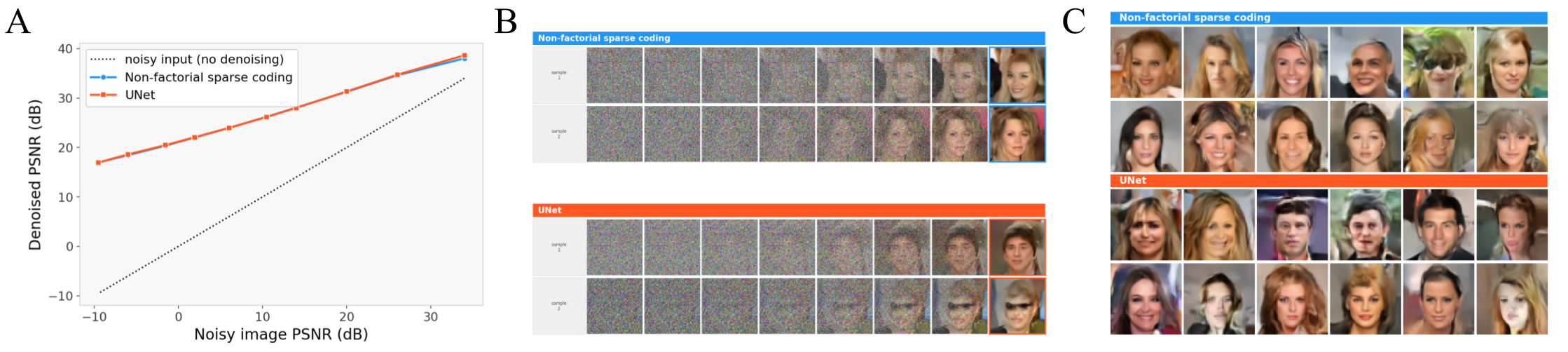}
\caption{{\bf Generation Performance.} (A) Denoising PSNR curves of our non-factorial sparse coding model versus a parameter-matched U-Net across varying noise levels. (B) Reverse diffusion sampling trajectories. (C) Final generated samples.}
    \label{fig:comparison}
\end{figure}


\label{result: comparison}

The nonfactorial sparse coding has consistently mirrored the complex generalization behaviors of opaque deep learning architectures. As previously shown, both this model and standard U-Nets \cite{UNET} exhibit identical contour completion capabilities (Figures~\ref{fig:intro}, \ref{fig:harmonic}) and develop semantic-level Geometry-Adaptive Harmonic Bases when trained on faces (Figure~\ref{fig:face}A). To confirm this functional equivalence extends to raw generative capability, we perform a final empirical sanity check against a parameter-matched U-Net (see detailed architecture in Appendix~\ref{app:unet}). Despite its bottom-up architectural simplicity, our model achieves  generative parity (with sampling algorithm described in Appendix~\ref{app:sampling}). Both models yield visually indistinguishable reverse diffusion samples and comparable denoising PSNRs across noise levels (Figure~\ref{fig:comparison}A-C). 
This confirms that the interpretable, biologically grounded architecture captures the core structural statistics of complex distributions just as effectively as standard black-box deep network-based models \cite{lecun2015deep, he2016deep, zeiler2014visualizing, zeiler2010deconvolutional}.

\section{Discussion}
\label{result: discussion}

We observed striking functional parallels between the visual cortex,  deep neural network (U-Nets), and a non-factorial sparse coding model. This convergence points toward a universal principle: these models are all capturing the same intrinsic statistics of natural images. The specific computations we discovered in our explicit model—such as short edges flexibly combining into continuous contours—are likely also occurring within black-box generative models as well. Growing evidence suggests that the black-box residual architectures underpinning standard diffusion models inherently perform a form of learned iterative, recurrent computation \cite{jastrzkebski2017residual, moon2024simple, gudovskiy2026odet}. Recognizing this structural equivalence offers a concrete new direction for mechanistic interpretability, such as new way to perform circuit analysis in large language model \cite{wang2022interpretability, lindseybatson}.

To fully understand this convergence, we must better characterize the structure of the natural image itself. We hypothesize that natural images act as sparse signals on a continuous manifold \cite{chen2018sparse, chen2022minimalistic,roweis2000nonlinear,tenenbaum2000global,he2026diffusion}. The manifold represents continuous, elementary structures (like edges), while the sparse signal dictates their composition. Approximating the distribution of natural images thus requires first parameterizing this manifold, and then estimating the density of these compositional functions. Sparse coding is interpretable precisely because it approximates this manifold in the simplest possible way: by dropping discrete "landmarks" onto it and interpolating the signal between them. It appears that visual cortex uses a similar strategy, likely driven by the metabolic necessity of sparsity \cite{atick1992does, field1994goal, karklin2011efficient}. We encourage both the computational neuroscience and deep learning communities to view their respective open challenges not as isolated domains, but as convergent solutions to the exact same geometric problem. Such insights are essential for building simpler, more efficient AI models and deepening our understanding of the brain.

\section*{Acknowledgments}
Z.Y. thanks Galen Chuang for helping reviewing and editing Section~\ref{V1 Hypotheses}. Z.Y. thanks David Lipshutz, Josue Casco-Rodriguez, Michael Fang, Ruichang Sun, Jack Kendall, Deepthi Bannai, Sam Reifenstein, Curtis McDonald, Weyl Lu, Eero Simoncelli, Chris Kymn, Galen Chuang, Wanyu Lei, Christina Savin, Anne Harrison, and Yury Polyachenko for helpful suggestions, discussions, and inspiration that led to this project. Z.Y. and B.A.O. were supported by the Center for the Co-Design of Cognitive Systems (CoCoSys), one of seven centers in JUMP 2.0, a Semiconductor Research Corporation (SRC) program sponsored by DARPA, and National Science Foundation Grant "Lie Group Representation Learning for Vision" (2313149). A.B. and B.A.O. are funded by the United States Air Force Office of Scientific Research (FA9550-21-1-0230).

\bibliographystyle{ieeetr}
\bibliography{references}

\newpage
\appendix
\renewcommand{\thefigure}{S\arabic{figure}}
\setcounter{figure}{0}
\section{Appendix}

\subsection{Derivation of score matching using latent variable model}
\label{app:score_matching}

Starting from the standard Denoising Score Matching objective, we substitute our MAP-approximated score (Equation \ref{eq:fix-point_approximate}) and the explicit noise formulation $x = y + \sigma \epsilon $, which means $\frac{\epsilon}{\sigma} = \frac{x - y}{\sigma^2}$:
\begin{align}
    \mathcal{L}_{\mathrm{DSM}}(\theta) &= \mathbb{E}_{\sigma, y, \epsilon} \left[ w(\sigma) \left\| \nabla_x \log p_{\theta,\sigma}(x) + \frac{\epsilon}{\sigma} \right\|_2^2 \right] \nonumber \\
    &\approx \mathbb{E}_{\sigma, y, \epsilon} \left[ w(\sigma) \left\| \frac{1}{\sigma^2}\big(\Phi z^*(x;\sigma,\theta) - x\big) + \frac{x - y}{\sigma^2} \right\|_2^2 \right] \nonumber \\
    &= \mathbb{E}_{\sigma, y, \epsilon} \left[ \frac{w(\sigma)}{\sigma^4} \big\| \Phi z^*(x;\sigma,\theta) - x + x - y \big\|_2^2 \right] \nonumber \\
    &= \mathbb{E}_{\sigma, y, \epsilon} \left[ \tilde{w}(\sigma) \left\| \Phi z^*(x;\sigma,\theta) - y \right\|_2^2 \right]
\end{align}
The noisy observation $x$ cancels algebraically. By defining the effective weighting schedule $\tilde{w}(\sigma) = \frac{w(\sigma)}{\sigma^4}$, this demonstrates that matching the score of the marginal data density perfectly reduces to minimizing the reconstruction error against the clean image $y$.

\textbf{Noise Distribution and Loss Weighting.} Having established the weighted reconstruction objective (Equation \ref{eq:dsm_to_weighted_recon_sigma4}), we must define the noise sampling distribution $p(\sigma)$ and the loss weighting schedule $w(\sigma)$. Following the preconditioning framework of Elucidating Diffusion Models (EDM) \cite{karras2022elucidating}, we sample the noise scale $\sigma$ from a log-normal distribution during training:
\begin{equation}
    \ln(\sigma) \sim \mathcal{N}(P_{\text{mean}}, P_{\text{std}}^2)
\end{equation}
where $P_{\text{mean}}$ and $P_{\text{std}}$ are hyperparameters governing the median and width of the noise distribution, respectively. This sampling strategy ensures the network focuses its capacity on the intermediate noise levels that are most critical for learning the data manifold, explicitly avoiding the extremes of pure noise or perfectly clean data.

To stabilize training and ensure that the expected gradient magnitudes remain uniform across all noise levels, we apply the EDM weighting schedule:
\begin{equation}
    w(\sigma) = \frac{\sigma^2 + \sigma_{\text{data}}^2}{(\sigma \cdot \sigma_{\text{data}})^2}
\end{equation}
where $\sigma_{\text{data}}$ is a hyperparameter representing the standard deviation of the underlying data distribution. By heavily penalizing errors at low noise levels (where $\sigma \to 0$) and down-weighting errors at high noise levels, this schedule dynamically balances the learning signal, allowing the optimal sparse coding dictionary $\Phi$ and latent state $z^*$ to be efficiently jointly optimized across all noise scales.


\subsection{Implicit gradient}
\label{app:implicit gradient}


Let $z^*(x;\sigma,\theta):=\lim_{t\to\infty} z_t(x;\sigma,\theta)$ denote the convergent state of the recurrent inference in Equation~\eqref{eq:ff_rec_update}, and let $\theta$ denote all model parameters, i.e., $\theta:=\{\Phi, M, \lambda, \gamma\}$. Both training the model under the denoising objective (Equation~\ref{eq:dsm_to_weighted_recon_sigma4}) and calculating the Jacobian of the score function $H(x)$ for analysis require differentiating through the convergent state $z^*(x;\sigma,\theta)$. Typically, computing gradients through a recurrent circuit requires Backpropagation Through Time (BPTT), a process that unrolls the entire iterative trajectory and incurs prohibitive computational and memory costs. However, because $z^*$ converges to a stable fixed point of the inference dynamics, we can completely bypass BPTT. The Implicit Function Theorem allows us to compute the exact analytical gradients in closed form: 
\begin{equation}
F(z; x,\sigma,\theta)
:=
z - T_{\theta,\sigma}(z;x),
\qquad
F(z^*; x,\sigma,\theta)=0.
\label{eq:fixed_point_residual}
\end{equation}
Assuming $z^*$ is an isolated stable fixed point and that $F$ is locally
differentiable with a non-singular Jacobian $\nabla_z F(z^*;x,\sigma,\theta)$,
the implicit function theorem yields
\begin{align}
\tfrac{\partial z^*}{\partial \theta}
&=
-\left[\nabla_z F(z^*;x,\sigma,\theta)\right]^{-1}
\nabla_\theta F(z^*;x,\sigma,\theta),
\label{eq:ift_grad1} \\
\tfrac{\partial z^*}{\partial x}
&=
-\left[\nabla_z F(z^*;x,\sigma,\theta)\right]^{-1}
\nabla_x F(z^*;x,\sigma,\theta).
\label{eq:ift_grad2}
\end{align}

We use the first implicit gradient $\tfrac{\partial z^*}{\partial \theta}$ to compute the exact gradient of model parameters for training, by chain rule: 
$\tfrac{d \mathcal{L}_{\mathrm{DSM}}}{d \theta} = \tfrac{\partial \mathcal{L}_{\mathrm{DSM}}}{\partial \theta} + \tfrac{\partial \mathcal{L}_{\mathrm{DSM}}}{\partial z^*} \tfrac{\partial z^*}{\partial \theta}.$ We utilize the second implicit gradient, $\tfrac{\partial z^*}{\partial x}$, to compute the exact analytical Jacobian of the denoiser. Because the denoised output is a linear function of the convergent state of inference, $f_\theta(x) = \Phi z^*(x; \sigma, \theta)$, applying the chain rule directly yields the denoising Jacobian: $J(x) = \Phi \frac{\partial z^*}{\partial x}.$

\subsection{Other training details}
\label{app:training tricks}

{\bf Exponential Moving Average (EMA) of Weights}
To stabilize the highly nonlinear, recurrent dynamics during training and to ensure smooth convergence of the generative prior, we maintain an exponential moving average (EMA) of the model weights. Rather than evaluating the network using the highly fluctuating active weights during inference, we use the EMA parameters. The decay rate $\beta$ is adjusted dynamically based on the global batch size ($N_{\text{batch}}$) and a predefined half-life ($k_{\text{img}}$) to ensure consistent smoothing independent of hardware scaling:

$$\beta = 0.5^{\frac{N_{\text{batch}}}{\max(k_{\text{img}} \times 1000, 1)}}$$

This practice is standard in state-of-the-art diffusion models, prevents mode collapse, and significantly improves the perceptual quality of the completed contours.

\subsection{Approximating Implicit Gradient}
\label{app:appendix_ift}
Given the implicit gradient for updating parameter as follows:
\begin{equation}
    \frac{\partial z^*}{\partial \theta} = -\left[\nabla_z F(z^*;x,\sigma,\theta)\right]^{-1} \nabla_\theta F(z^*;x,\sigma,\theta),
\end{equation}
Computing the matrix inverse of the fix point update's dynamics $\left[F(z^*;x,\sigma,\theta)\right]^{-1}$ is expensive, and computationally intractable if $z^*$ is extremely high dimension like convolution sparse code for images. In practice, we approximate $\left[F(z^*;x,\sigma,\theta)\right]^{-1}$. Recall we define 
$$
F(z; x,\sigma,\theta)
:=
z - T_{\theta,\sigma}(z;x),
\qquad
F(z^*; x,\sigma,\theta)=0.
$$
So the Jacobian of $F$ is $[I-J]^{-1}$, where $J$ is the Jacobian of single step ISTA update $T_{\theta,\sigma}(z;x)$.  Assuming the ISTA update is stable (the spectral radius of $J$ is strictly less than 1), then we can approximate $[I-J]^{-1}$  using truncated neumann-series expansion: $\sum_{k=0}^{m} J^k$. While this Neumann-series Phantom Gradient (NPG) can be explicitly computed via iterative vector-Jacobian products, it is mathematically equivalent to simply unrolling the forward dynamics. Therefore, following the method derived in \cite{geng2021training,fung2022jfb}, we implement an Unrolling-based Phantom Gradient (UPG): after rapidly finding the detached fixed point $z^*$ without gradient tracking, we unroll the dynamics from $z^*$ for $m$ additional iterations with gradients enabled to compute $z_m = T^m(z^*; x, \sigma, \theta)$. During backpropagation, evaluating $\frac{\partial z_m}{\partial \theta}$ via the standard chain rule naturally accumulates the exact Neumann series expansion. We generally set $m$ between 1 and 3 steps during training. This allows us to efficiently approximate the implicit gradient by leveraging standard autograd engines, bypassing the severe memory constraints of full backpropagation through time.

\subsection{Detailed intuition on the mechanistic decomposition of the denoising Jacobian}
We provide a more detailed interpretation on the decomposition of the denoising Jacobian in Equation~\ref{eq:jacobian-change-basis} and Equation\ref{eq:jacobian-decomposition}:

\begin{equation*}
J(x) = \frac{d x_{\text{clean}}}{d x_{\text{noise}}} = \Phi \underbrace{\eta \left[ I - \Sigma(I - \eta W) \right]^{-1} \Sigma}_{J_z} \Phi^\top = \Phi J_z \Phi^\top
\end{equation*}
\begin{equation*}
    J_{z,i} = \underbrace{e_i}_{\text{Sparse gating}} + \underbrace{\eta \Sigma W e_i}_{\text{1st-order spread}} + \underbrace{\eta^2 \Sigma W \Sigma W e_i}_{\text{2nd-order spread}} + \dots \quad (\text{if } z^*_i > 0, \text{ else } 0)
\end{equation*}

\begin{enumerate}

    \item \textbf{Projection and Inhibition ($\Sigma \Phi^\top$):} The perturbed input is projected onto the dictionary. The $\Sigma$ mask instantly silences inactive elements, ensuring that only perturbations relevant to the current context survive, analogous to coring or hard thresholding in signal processing. 

    \item \textbf{Context-Aware Spread ($\Sigma W$):} Surviving signals spread via lateral connections $W$. The outer $\Sigma$ clamps this expansion, isolating local geometric cross-talk by ensuring perturbations only spread to activated neurons. 

    \item \textbf{Higher-Order Context-Aware Spread $(\Sigma W)^k$:} The infinite tail of the sum temporally unrolls the dynamics. Just as raising an adjacency matrix to the $k$-th power computes $k$-hop context-aware spread of activity defined by lateral connection. Higher-order term $(\Sigma W)^k$ allows the neural activity not only spread but ricochets globally yet strictly confined among mutually active neurons. 

    \item \textbf{Synthesis ($\Phi$):} The context-modulated signals are projected back into ambient pixel space.
\end{enumerate}

\subsection{Sampling algorithm}
\label{app:sampling}

Standard diffusion models generate samples by simulating a stochastic differential equation (SDE) in pixel space, such as the variance-preserving (VP) SDE:
\begin{equation}
    d x_t = \Big[-\tfrac{1}{2}\beta(t) x_t - \beta(t)\,\nabla_{x}\log p_t(x_t)\Big]\,dt + \sqrt{\beta(t)}\, dW_t
    \label{eq:vp_sde}
\end{equation}
where $\nabla_{x}\log p_t(x_t)$ is the score function of the perturbed data distribution. 

Following the deterministic formulation of \cite{karras2022elucidating}, we can map this generation process to a Probability Flow Ordinary Differential Equation (ODE). By parameterizing the time variable directly by the noise standard deviation $\sigma$, the deterministic ODE governing the data evolution is given by:
\begin{equation}
    \mathrm{d}x = -\sigma \nabla_x \log p_{\theta,\sigma}(x) \, \mathrm{d}\sigma
    \label{eq:score_ode}
\end{equation}
In practice, the score is parameterized via a denoiser model $D_\theta(x, \sigma)$, which relates to the score via Tweedie's formula, $D_\theta(x, \sigma) = x + \sigma^2 \nabla_x \log p_{\theta,\sigma}(x)$. Substituting this relationship into Equation~\ref{eq:score_ode} yields the ODE in terms of the denoiser:
\begin{equation}
    \mathrm{d}x = \frac{x - D_\theta(x, \sigma)}{\sigma} \, \mathrm{d}\sigma
    \label{eq:edm_ode}
\end{equation}
In our framework, evaluating the denoiser $D_\theta(x, \sigma)$ is parametrize as unrolling the MAP estimate recurrent dynamics to approximate the fixed point, i.e., $D_\theta(x, \sigma) = \Phi z_K(x, \sigma)$.

To numerically integrate Equation~\ref{eq:edm_ode}, we employ a second-order Heun (improved Euler) scheme. The continuous noise level $\sigma$ is discretized into a monotonically descending sequence $\{\sigma_0, \sigma_1, \dots, \sigma_N\}$, starting from pure noise $\sigma_0 = \sigma_{\max}$ and terminating at a minimum noise level $\sigma_N = \sigma_{\min}$. At each integration step $i$, the solver performs a predictor-corrector update. First, an Euler step predicts the intermediate state at $\sigma_{i+1}$. Then, a corrector step evaluates the denoiser at this predicted state. The final update is computed using the trapezoidal rule by averaging the derivatives. The full sampling procedure is summarized in Algorithm~\ref{alg:heun}.

\begin{algorithm}[h]
\caption{Annealed Heun Sampling}
\label{alg:heun}
\begin{algorithmic}[1]
\REQUIRE Denoiser $D_\theta$, recurrent iterations $K$, noise schedule $\{\sigma_0, \sigma_1, \dots, \sigma_N\}$
\STATE Sample $x_0 \sim \mathcal{N}(0, \sigma_0^2 \mathbf{I})$
\FOR{$i = 0$ \TO $N-1$}
    \STATE $\hat{x}_0 \leftarrow D_\theta(x_i, \sigma_i)$ \COMMENT{Evaluated as $\Phi z^*(x_i,\sigma_i)$}
    \STATE $d_i \leftarrow \frac{x_i - \hat{x}_0}{\sigma_i}$
    \STATE $x_{\text{euler}} \leftarrow x_i + (\sigma_{i+1} - \sigma_i) d_i$
    \vspace{0.1cm}
    \STATE \textit{\% Corrector Step}
    \STATE $\hat{x}_1 \leftarrow D_\theta(x_{\text{euler}}, \sigma_{i+1})$ \COMMENT{Evaluated as $\Phi z^*(x_{\text{euler}}, \sigma_{i+1})$}
    \STATE $d'_{i+1} \leftarrow \frac{x_{\text{euler}} - \hat{x}_1}{\sigma_{i+1}}$
    \vspace{0.1cm}
    \STATE \textit{\% Heun Update}
    \STATE $x_{i+1} \leftarrow x_i + (\sigma_{i+1} - \sigma_i) \left( \frac{1}{2} d_i + \frac{1}{2} d'_{i+1} \right)$
\ENDFOR
\STATE $x_{\text{final}} \leftarrow D_\theta(x_N, \sigma_N)$ \COMMENT{Final projection at $\sigma_{\min}$}
\RETURN $x_{\text{final}}$
\end{algorithmic}
\end{algorithm}

\subsection{Negative Log likelihood Estimation}
\label{appendix:BPD}
For our model, which directly predicts the clean data point $x_0$ via a parameterized denoising network $D_\theta(x, \sigma)$, the probability flow ODE with respect to the continuous noise scale $\sigma$ takes the following form:$$\frac{\mathrm{d}x}{\mathrm{d}\sigma} = \frac{x - D_\theta(x, \sigma)}{\sigma} =: \tilde{f}_\theta(x, \sigma)$$Using the instantaneous change of variables formula, we can compute the marginal log-likelihood of the data at the minimum noise level $\sigma_{\text{min}}$ by integrating the divergence of the empirical drift function up to the maximum noise level $\sigma_{\text{max}}$:$$\log p_{\sigma_{\text{min}}}(x(\sigma_{\text{min}})) = \log p_{\sigma_{\text{max}}}(x(\sigma_{\text{max}})) + \int_{\sigma_{\text{min}}}^{\sigma_{\text{max}}} \nabla_x \cdot \tilde{f}_\theta(x(\sigma), \sigma) \mathrm{d}\sigma$$where the random variable $x(\sigma)$ as a function of $\sigma$ is obtained by solving the probability flow ODE. In high-dimensional spaces, exactly computing the divergence $\nabla_x \cdot \tilde{f}_\theta(x, \sigma)$ is computationally prohibitive. Following Grathwohl et al. (2018), we approximate it using the unbiased Skilling-Hutchinson trace estimator:
$$\nabla_x \cdot \tilde{f}_\theta(x, \sigma) = \mathbb{E}_{p(\epsilon)} [\epsilon^\top \nabla_x \tilde{f}_\theta(x, \sigma) \epsilon]$$where $\nabla_x \tilde{f}_\theta$ denotes the Jacobian of $\tilde{f}_\theta(\cdot, \sigma)$, and the random projection vector $\epsilon \sim \mathcal{N}(0, \mathbf{I})$. The vector-Jacobian product $\epsilon^\top \nabla_x \tilde{f}_\theta(x, \sigma)$ can be efficiently evaluated in a single backward pass using reverse-mode automatic differentiation. To accelerate the evaluation of the trace integral, we solve the probability flow ODE using a first-order Euler method over a geometric (log-linear) schedule of $N=20$ discrete steps from $\sigma_{\text{min}}$ to $\sigma_{\text{max}}$. We assume the prior distribution at the boundary condition $\sigma_{\text{max}}$ is a standard isotropic Gaussian, yielding $p_{\sigma_{\text{max}}}(x) = \mathcal{N}(0, \sigma_{\text{max}}^2 \mathbf{I})$. Finally, to report the likelihood results, we convert the computed continuous log-likelihoods to discrete Bits Per Dimension (BPD). We divide the log-likelihood by $D \ln(2)$ (where $D$ is the data dimensionality) and apply an 8.0 bit dequantization offset to account for the uniform binning of standard 8-bit [0, 1] scaled image data.

\subsection{Practical implemention of model's dynamics: Multi-Scale Convolutional energy function, and Asymmetric update.}
While Equation \ref{eq:ff_rec_update} provides a theoretically exact formulation for a single-scale, flatten vector space, adapting this inference dynamics to high-dimensional natural images requires a multi-scale, convolutional architecture. In our practical implementation, the dictionary $\Phi$ and the prior interaction matrix $M$ are distributed across multiple spatial scales (akin to a Gaussian pyramid) to capture both fine local edges and coarse global contours. Furthermore, we employ an asymmetric, alternating update schedule to accelerate the propagation of information across the network.

{\bf Multi-Scale Likelihood via Laplacian Pyramid} 
Instead of a single dense dictionary, the network utilizes a set of scale-specific convolutional dictionaries \cite{chalasani2013fast} $\Phi^{(i)}$ for scales $i = 0, \dots, L-1$ (where $i=0$ is the finest spatial resolution). To map these latent variables to the image space, we define $P$ as the multiscale Laplacian pyramid synthesis operator. The multiscale reconstruction is formulated as $P(\Phi z)$, allowing us to rewrite the joint energy function's likelihood term as:

$$E_{\text{likelihood}} = \frac{1}{2 \sigma^2} \|x - P \Phi z\|_2^2$$

To optimize this, the gradient with respect to $z$ requires applying the chain rule, yielding $-\Phi^\top P^\top (x - P \Phi z)$. Here, $P^\top$ acts as the Laplacian pyramid analysis operator, decomposing the image-space residual into scale-specific frequency bands. The feedforward drive for a specific scale $i$ is then computed by passing this banded residual through the corresponding transposed dictionary:

$$\text{Feedforward Drive}^{(i)} = \Phi^{(i)\top} \big[ P^\top (x - P \Phi z) \big]^{(i)}$$

\begin{figure}
    \centering
\includegraphics[width=0.5\linewidth]{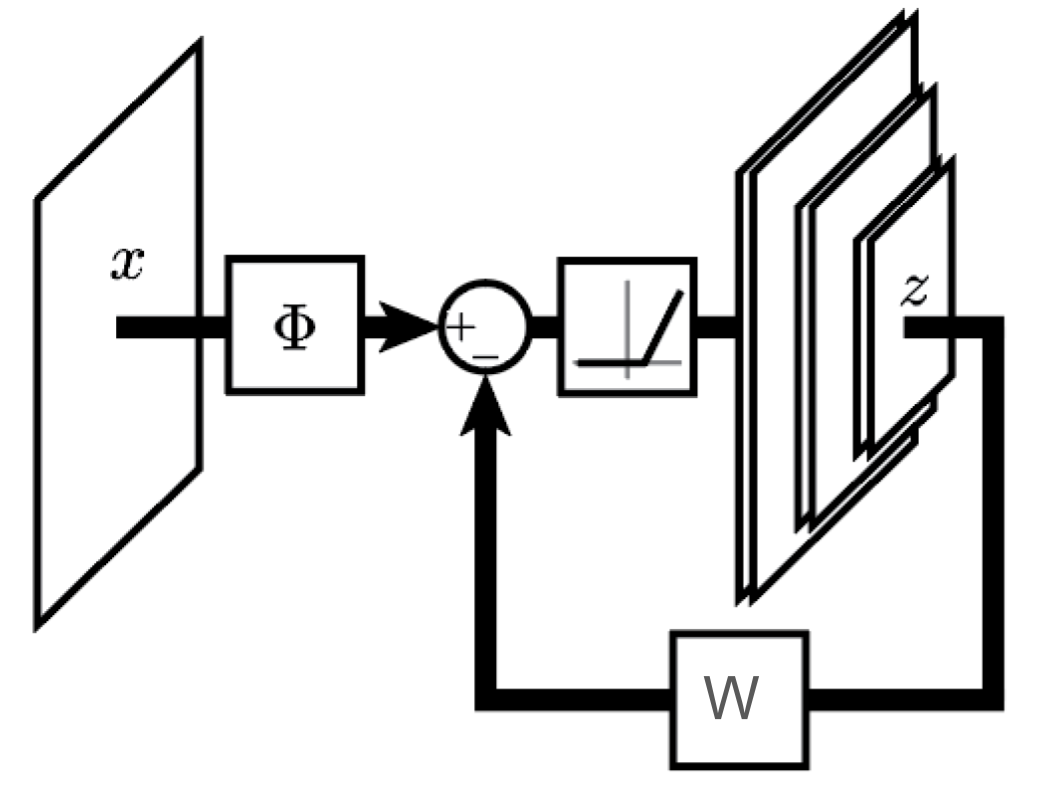}
\caption{Schematic of the multiscale, convolutional sparse coding's inference dynamics.}
    \label{fig:schematic}
\end{figure}

To explicitly model the non-factorial co-occurrence statistics across different spatial scale, the global prior matrix $M$ is defined as a block-tridiagonal operator. This restricts interactions strictly to intra-scale horizontal connections and adjacent inter-scale vertical connections. We can thus decompose the gradient of $z^T M z$ co-occurrence part the energy term wrt $z$ in the following way:

For any scale $i$, the prior drive is composed of three distinct convolutional operators: 
\begin{enumerate}
    \item Lateral Scale Connections ($W_{\text{lat}}^{(i)}$): Captures intra-scale horizontal co-occurrences (e.g., collinear edges).
    \item  Up-Scale Connections ($W_{\text{up}}^{(i)}$): Routes feedforward contextual signals from the finer, higher-resolution scale $i-1$.
    \item Up-Down Connections ($W_{\text{down}}^{(i)}$): Routes feedback signals from the coarser, lower-resolution scale $i+1$. 
\end{enumerate}
which results in the following ISTA inference update step for multiscale version version of non-factorial sparse coding:

\begin{equation}
z_{t+1}^{(i)} = \text{ReLU} \left( z_t^{(i)} + \eta \left[ \underbrace{\Phi^{(i)\top} \big[ P^\top r_t \big]^{(i)}}_{\text{Feedforward Drive}} + \sigma^2\gamma(\sigma) \underbrace{\left( W_{\text{lat}}^{(i)} z_t^{(i)} + W_{\text{up}}^{(i)} z_t^{(i-1)} + W_{\text{down}}^{(i)} z_t^{(i+1)} \right)}_{\text{Hierarchical Recurrent Prior}} - \underbrace{\gamma(\sigma)\lambda^{(i)}}_{\text{Sparsity Bias}} \right] \right)
\end{equation}

where $r_t = x - P \Phi z_t$ is the current global image-space reconstruction residual. In addition, all the matrix multiplication in this model is implemented using convolution, and transpose of the matrix is implemented using transposed convolution, as illustrated Figure~\ref{fig:schematic}.

{\bf Asymmetric (Sweeping) Updates} In a standard, synchronous ISTA implementation (Jacobi-style updates), information strictly propagates one scale per iteration, creating a latency of $L$ steps for the finest scale to influence the coarsest scale. To overcome this and simulate rapid biological inference, our implementation utilizes an asymmetric sweeping update schedule (Gauss-Seidel optimization). A single inference iteration consists of a continuous two-pass sweep: Bottom-Up Sweep: The network updates scales sequentially from finest to coarsest ($i = 0 \to L-1$). During this pass, the calculation for $z^{(i)}$ utilizes the already updated values from $z_{t+1}^{(i-1)}$, allowing fine-grained local evidence to immediately percolate up the hierarchy. Top-Down Sweep: The network reverses direction, updating from coarsest to finest ($i = L-1 \to 0$). Here, the calculation for $z^{(i)}$ utilizes the already updated coarse context from $z_{t+1}^{(i+1)}$, rapidly broadcasting global shape constraints down to the local feature detectors. Empirically, this update rule help both model to get to fix point faster and parameter convergence faster.

\newpage
\begin{figure}[t]
    \centering
\includegraphics[width=0.85\linewidth]{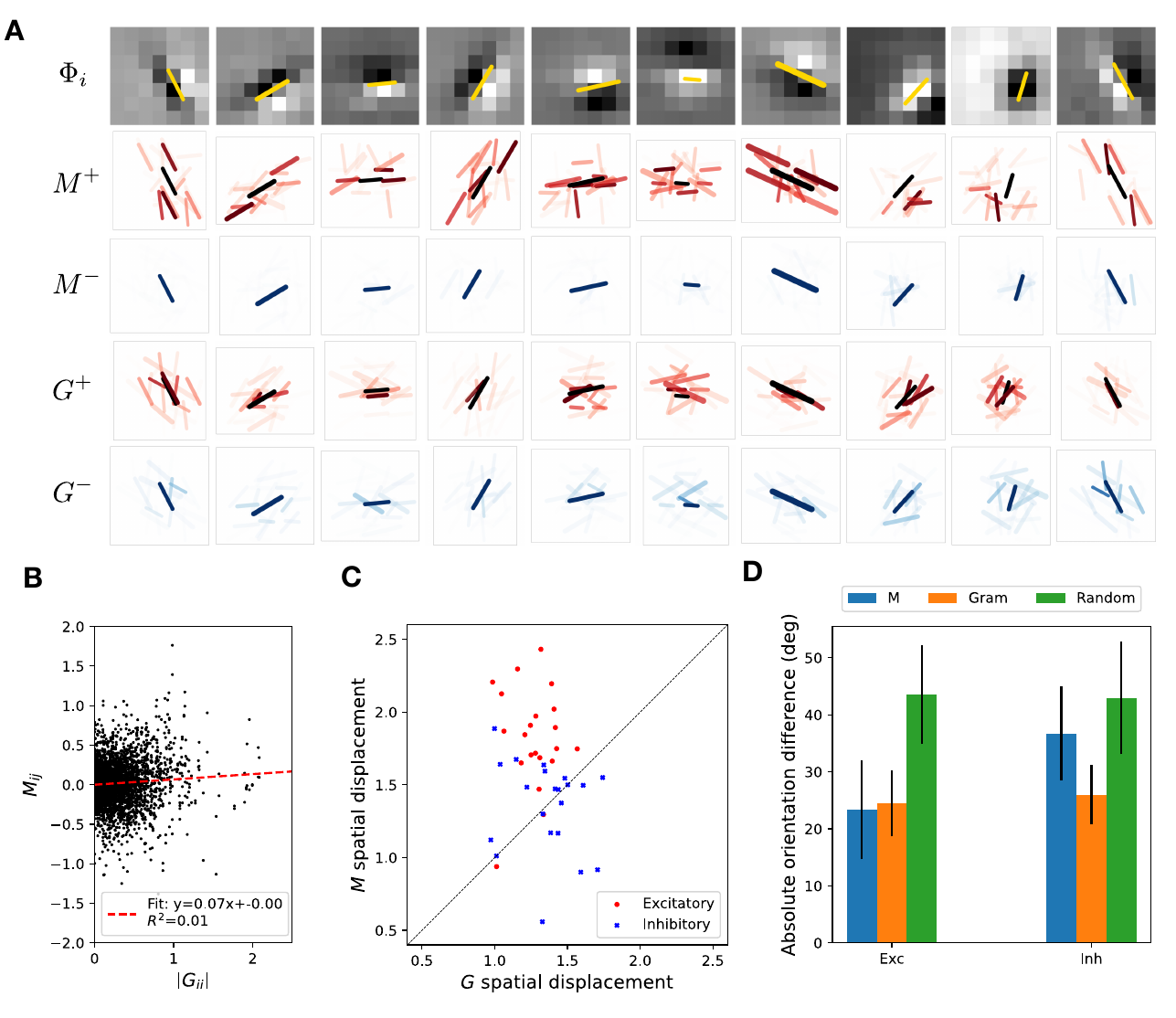}
    \caption{
        \textbf{(A)} 
        Additional learned interaction maps using the same needle-plotting convention as Figure~\ref{fig:neuroscience}B. 
        Columns correspond to reference basis functions $\Phi_i$ and colored bars show interactions with other basis functions $\Phi_j$.
        \text{(B)}
        Joint scatter plot comparing learned pairwise interactions $M_{ij}$ with Gram interactions $G_{ij}=\Phi_i^\top\Phi_j$. 
        The red dashed line shows the linear fit. Its positive slope indicates that learned interactions are biased toward facilitation
        \textbf{(C)}
        Spatial displacement of the top-10 largest-magnitude interactions for each neuron, comparing learned interactions $M$ with Gram interactions $G$ separately for excitatory and inhibitory components. 
        Learned excitatory interactions span larger spatial displacements than Gram interactions, indicating longer-range facilitation. A less significant difference  
        \textbf{(D)}
        Average absolute orientation difference between neurons and its top-10 connected units for $M$, $G$, and randomly assigned connections. Error bars indicate one standard deviation. 
        Learned excitatory interactions connect units with similar preferred orientations, whereas learned inhibitory interactions show weaker orientation similarity.
    }
    \label{fig:h_conn_summary}
\end{figure}

\subsection{More visualization eigenvectors of Jacobian and decomposition}
\begin{figure}
    \centering
\includegraphics[width=\linewidth]{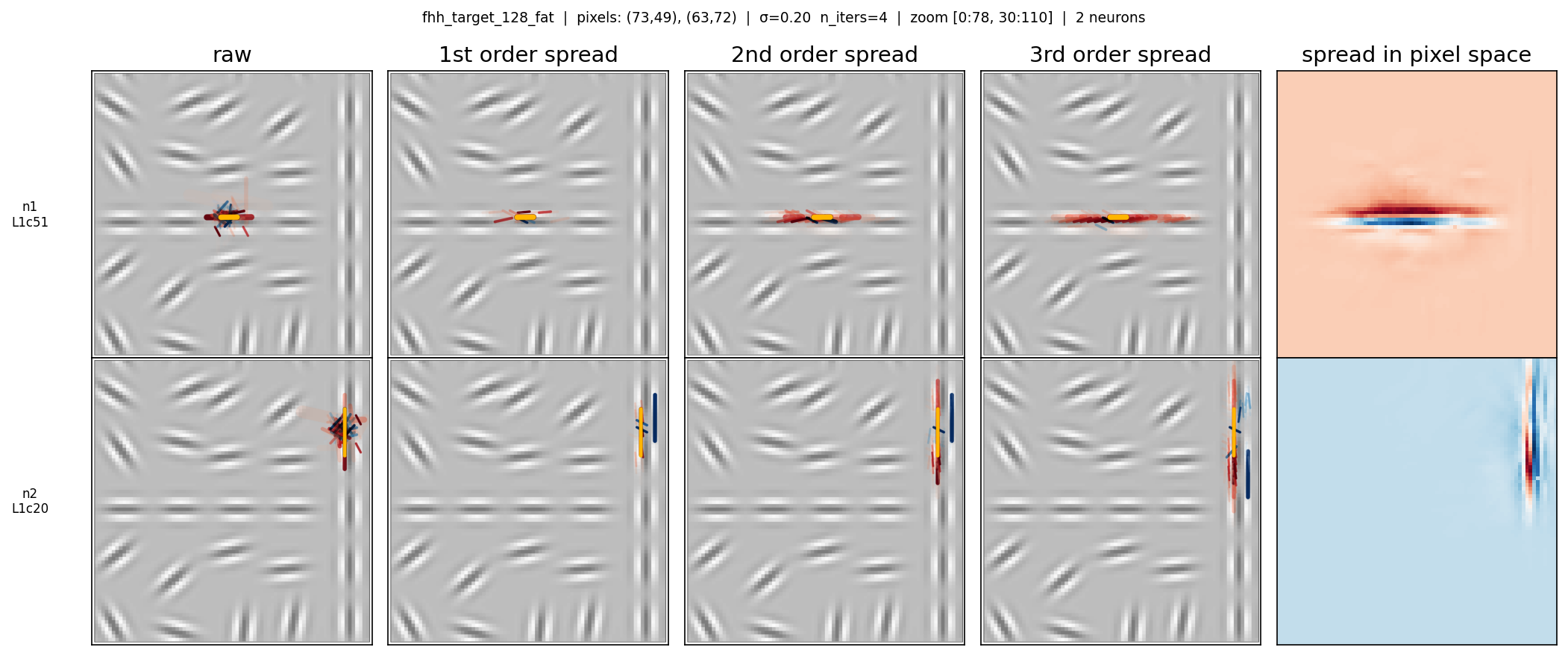}
\caption{More latent Jacobian impulse decomposition visualized in Figure~\ref{fig:harmonic}.D.}
    \label{fig:more_jacobian_spread}
\end{figure}

See Figure~\ref{fig:more_jacobian_spread}.

\begin{figure}
    \centering
\includegraphics[width=0.8\linewidth]{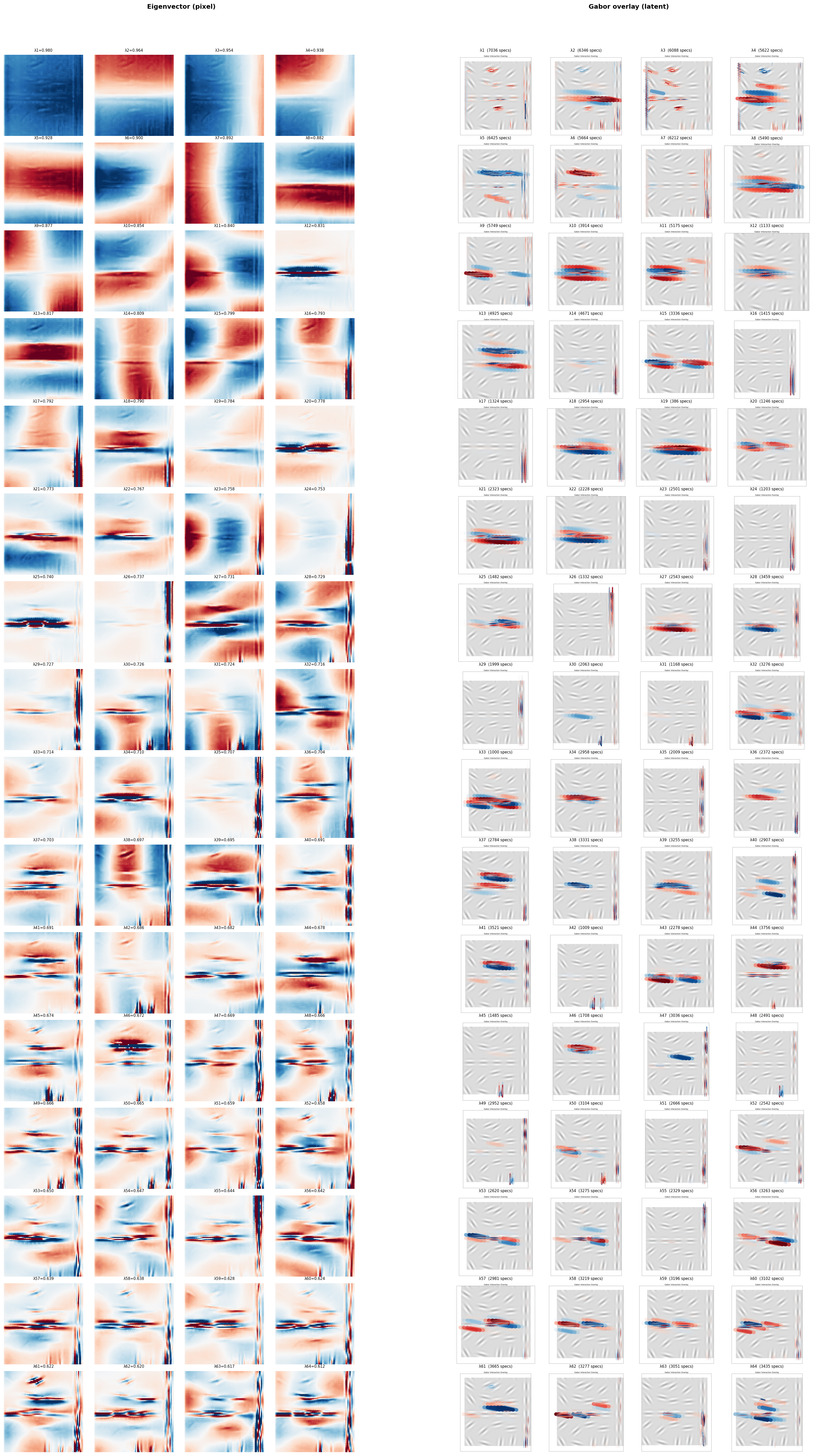}
\caption{More pair of eigenvectors of pixel vs latent Jacobian of non-factorial sparse coding model visualized in Figure~\ref{fig:harmonic}.A.}
    \label{fig:more_eigenvis}
\end{figure}

\subsection{U-Net Architecture Details}
\label{app:unet}

For our experiments, we employ a  U-Net architecture based on the guided-diffusion implementation \cite{dhariwal2021diffusion}. Specifically, we use the following configuration for processing $64 \times 64$ resolution images: Base Architecture: The model takes 3-channel inputs and outputs 3 channels, starting with a base feature width of $32$ channels. Depth and Channel Scaling: The network consists of 4 resolution levels with channel multipliers of $(1, 2, 3, 4)$. This results in feature map widths of $32$, $64$, $96$, and $128$ at each respective stage of the encoder and decoder. Residual Blocks: Each resolution level contains $1$ residual block. Downsampling and upsampling operations are also parameterized as residual blocks. We apply dropout with a probability of $0.1$ within the residual blocks, and we do not use scale-shift normalization. Attention: To maintain a fully convolutional, minimal architectural footprint, all spatial self-attention blocks are disabled. Timestep Conditioning: The noise scale (or timestep) is projected into a 64-dimensional embedding space (using a factor of $2$ relative to the base width) and is injected into every residual block via standard addition.

\end{document}